%% file: FNAL_preprint.tex
\RequirePackage{lineno}

\documentclass[aps,prl,twocolumn,showpacs,superscriptaddress]{revtex4} 

\usepackage{amssymb}
\usepackage{amsmath}
\usepackage{graphicx}
\usepackage{dcolumn}
\usepackage{epsfig}

\newcommand{\met}       {\mbox{$\not\!\!p_T$}}

\newcommand{\ppbar}     {\mbox{$p\bar p$}}
\newcommand{\Wbb}       {\mbox{$W\!+b\bar b$}}
\newcommand{\Wcc}       {\mbox{$W\!+c\bar c$}}

\newcommand{\Zbb}       {\mbox{$Z\!+b\bar b$}}
\newcommand{\Zcc}       {\mbox{$Z\!+c\bar c$}}

\newcommand{\ifb}{\,\mbox{fb}^{-1}}
\newcommand{\pb}{\,\mbox{pb}}

\newcommand{\TeV}{\,\mbox{Te\kern-0.2exV}}
\newcommand{\GeV}{\,\mbox{Ge\kern-0.2exV}}
\newcommand{\mGeV}{\,\mathrm{Ge\kern-0.2exV}}
\newcommand{\MeV}{\,\mbox{Me\kern-0.2exV}}
\newcommand{\keV}{\,\mbox{ke\kern-0.2exV}}
\newcommand{\eV}{\,\mbox{e\kern-0.2exV}}

\begin{document}
\setpagewiselinenumbers
\hspace{5.2in} \mbox{FERMILAB-PUB-05/599-E}
\title{\boldmath
Search for a Narrow ${t\bar{t}}$ Resonance in ${p\bar{p}}$ Collisions at $\sqrt{s}=1.96$~TeV}
\vspace*{2.cm}

\input author_list.tex

\date{November 4, 2011}
           

\begin{abstract}
We report a search for a narrow $t\bar{t}$ resonance that decays into a lepton+jets 
final state based on an integrated luminosity of $5.3$~fb$^{-1}$ of proton-antiproton collisions at $\sqrt{s}=1.96$~TeV collected by the D0 Collaboration at the Fermilab Tevatron Collider. 
We set upper limits on the production cross section of such a resonance multiplied by its 
branching fraction to $t\bar{t}$ which we compare to predictions for a leptophobic topcolor $Z'$
boson. We exclude such a resonance at the 95\%~confidence level for masses below $835$~GeV. 
\end{abstract}

\pacs{14.65.Jk, 13.85.Rm}

\maketitle

\newpage
Narrow resonances that decay to top-antitop quark ($t\bar t$) pairs
are predicted by many models of physics beyond the standard model. 
Such heavy neutral gauge bosons ($X$) appear in grand unified theories with
symmetry groups larger than SU(5)~\cite{Leike:1998wr}, as Kaluza-Klein excitations of
the gluon or of the $Z$ boson that predominantly decay to $t\bar{t}$ pairs~\cite{Lillie:2007yh,Rizzo:1999en},
as axigluons~\cite{Sehgal:1987wi}, and in theories of new strong
dynamics~\cite{Hill:1993hs,Harris:1999ya}. Some resonances are expected to
have a large decay width and would be hard to 
detect over standard model $t\bar{t}$-pair production. Here we
search for a $t\bar t$ resonance with a width that is significantly smaller than the detector resolution for reconstructing its mass. 

Searches for $t\bar t$ resonances were previously carried out
by the D0 and CDF collaborations at the Fermilab Tevatron Collider, and no
evidence was found for resonant production~\cite{Abazov:2008ny,Aaltonen:2011ts}. 
To extract limits on the mass $M_X$ of a resonance,
limits on cross sections were compared to predictions for a leptophobic topcolor 
$Z'$ boson with width $\Gamma=0.012 M_{Z'}$~\cite{Harris:1999ya}. 
To simplify comparisons with previous limits, we use the same model
as a reference. The lower limit for the mass of such a topcolor $Z'$ boson published
by the D0 Collaboration is $700\GeV$~\cite{Abazov:2008ny} based on $0.9\ifb$.
The CDF Collaboration excludes $Z'$ boson masses below $900\GeV$~\cite{Aaltonen:2011ts} 
using $4.8\ifb$ of integrated luminosity. Both limits correspond to $95\%$~confidence level (C.L.).

We carry out the search using lepton+jets ($\ell+$jets) events in which one of the top
quarks decays  ``leptonically'', $t\rightarrow Wb\rightarrow \ell\bar\nu b$,
 and the other ``hadronically'', $t\rightarrow Wb\rightarrow q\bar q'b$ (charge conjugate states are implicitly included in our notations). 
This final state is characterized by an isolated charged
lepton (electron or muon), imbalance in  transverse momentum $\met$ from the
undetected neutrino, and jets from the fragmentation of the four quarks. 
The data correspond to an integrated luminosity of $5.3\ifb$ and 
were acquired by the D0 experiment at the Fermilab Tevatron Collider
in $p\bar{p}$ collisions at $\sqrt{s}=1.96\TeV$. 

The D0 detector consists of central tracking, calorimeter, and muon
systems~\cite{Abachi:1993em,Abazov:2005pn}. The central tracking system is located inside
a $1.9\,$T superconducting solenoidal magnet. Central and forward preshower
detectors are located just outside the coil and in front of the
calorimeters. The liquid-argon/uranium sampling calorimeter is divided into a central
section covering pseudorapidity $|\eta|<1.1$ and two end calorimeters
extending coverage up to $|\eta|\approx4$. The calorimeter is segmented
longitudinally into electromagnetic, fine hadronic, and coarse hadronic
sections with increasingly coarser granularity. The muon system, located outside
the calorimeter, consists of one layer of tracking detectors and scintillation
trigger counters inside a $1.8\,\mbox{T}$ toroidal magnet and two similar layers outside
the toroids. A three-level trigger system selects events that are recorded for
offline analysis. 

Events must satisfy one of several trigger conditions, all requiring an electron or muon with high transverse momentum, in some cases in conjunction with one or more jets. The event selection requires one isolated lepton with $p_T>20$~GeV, missing
transverse momentum above 20~GeV (30~GeV) for the $e$+jets ($\mu$+jets) data,
and at least three jets with $p_T>20$~GeV. The leading jet must have $p_T>40$~GeV. 
We require at least one jet to be
tagged as originating from the fragmentation of a $b$~quark. Further details 
about the $\ell$+jets event selection can be found in \cite{Abazov:2011_tprime} 
and \cite{Abazov:2011mi}.
After applying these criteria, the dominant background is continuum $t\bar{t}$ 
production. We discriminate between continuum and resonant $t\bar t$ production 
using the invariant mass of the $t\bar{t}$ system.  
The $\ell+$jets events are divided into four subsamples defined by lepton flavor 
($e$, $\mu$) and jet multiplicity (3 jets and $\geq4$~jets).

The two main standard model processes that yield an isolated 
lepton, $\met$, and several jets are 
$t\bar{t}$ and $W$+jets production. The third most important background 
arises from multijet events in which a jet is misidentified
as an electron, or a muon from heavy-flavor quark decay appears isolated, and $\met$ is mismeasured. Single top
quark, $Z$+jets, and diboson production can also give rise to such final
states, but have much smaller yields.

We simulate $t\bar t$, $W$+jets and
$Z$+jets production using the {\sc alpgen+pythia} 
event generators~\cite{Mangano:2002ea,Sjostrand:2006za,Hoche:2006ph}.
Three subsamples, \Wbb, \Wcc, and $W\!+$light-partons, are generated separately. 
Similarly, the $Z$+jets samples are divided into \Zbb, \Zcc, and $Z+$light-parton samples.
We simulate single top quark production using the {\sc comphep}-{\sc
  singletop}~\cite{Boos:2006af,Boos:2004kh} generator and diboson 
($WW$, $WZ$, and $ZZ$) production with {\sc pythia}. For all simulations, we set the top
quark mass to $m_t=172.5\GeV$ and use the CTEQ6L1 parton distribution
functions~\cite{Pumplin:2002vw}. We simulate detector effects 
using {\sc  geant-3}~\cite{Brun:1993} 
and add randomly triggered events to all
simulated events to account for multiple \ppbar\ collisions in the same bunch
crossing. These events are 
reconstructed using the same procedures as for data. 

To estimate backgrounds we use either data-driven methods or simulation. We estimate the number and distribution of multijet events that are contained in
each subsample using a control data sample~\cite{Abazov:2007kg}.
We scale $t\bar t$ production to the theoretical approximate next-to-next-to-leading order
(NNLO) prediction of $\sigma(p\bar{p}\rightarrow t\bar{t})=7.48^{+0.56}_{-0.72}\pb$~\cite{Moch:2008qy}. 
We normalize single top quark production in the $s$ and $t$~channels to the
NNLO cross section with next-to-NNLO threshold corrections of $3.3\pb$~\cite{Kidonakis:2008mu}. 
We normalize diboson processes to their next-to-leading order (NLO) cross sections, as computed with {\sc mcfm}~\cite{VVxsec}, of $12.0\pb$ for
$WW$, $3.7\pb$ for $WZ$, and $1.4\pb$ for $ZZ$ production.
We fix the relative normalization of the $Z+b\bar b$, $Z+c\bar c$, and $Z$+light 
parton samples and the relative normalizations of the \Wbb, \Wcc, and $W$+light parton
samples to NLO predictions computed with {\sc mcfm}~\cite{VVxsec}. 
We then normalize the inclusive $Z$ boson production such that $\sigma(p\bar{p}\rightarrow Z)\times B(Z\rightarrow \mu^+\mu^-)$ agrees with the NNLO prediction of 
$256\pb$~\cite{Hamberg:1990np}.
We normalize the number of $W$+jets events such that the total number 
of events predicted by all background sources equals the number of events observed in each 
of the four subsamples before imposing the $b$-tagging requirement. 
This corresponds to increasing the total number of $W$+jets events expected by a factor of approximately $1.3$ which is close to the NLO k-factor for $W$+jets production.

As a model for $t\bar{t}$ resonance production, we use production of a 
$Z'$ boson in {\sc pythia}, that decays exclusively to $t\bar{t}$ pairs.
We consider 18 resonance mass values $M_X$ between $350$ and $1200\GeV$.
The intrinsic widths of the resonances are set to $\Gamma_X=0.012M_X$.

We reconstruct the $t\bar{t}$ invariant mass, $m_{t\bar{t}}$, using up to four jets with the 
highest $p_T$, the charged lepton, and the neutrino. We determine the momentum 
of the neutrino by equating the neutrino $p_T$ to the measured $\met$
constraining the invariant mass of the charged lepton-neutrino system to
the $W$ boson mass and choosing the smaller solution of the resulting quadratic equation 
for the neutrino momentum component $p_z$ along the beam direction. If there is 
no real solution we set neutrino $p_z=0$ and scale the $\met$ to satisfy the $W$ boson mass constraint. Figure~\ref{fig:MCdist} shows the expected distribution of
the reconstructed $m_{t\bar{t}}$ from the production of narrow
$t\bar{t}$ resonances of  different mass values.
The detector resolution for $m_{t\bar t}$ varies between $65\GeV$ for $M_X=400\GeV$ 
and $270\GeV$ for $M_X=1.2\TeV$, which is much larger than the widths of the $t\bar t$ 
resonances considered.

We use the reconstructed $t\bar{t}$ mass to test for the presence of a signal in the data and to compute upper limits on the production cross section of a narrow $t\bar{t}$ resonance times branching fraction to $t\bar t$, $\sigma B$, as a function of its mass. For each hypothesized value of the resonance mass, we fit the data to background-only and to signal+background hypotheses. For each hypothesis we also vary the systematic uncertainties given in Table~\ref{tab:syst} subject to a Gaussian constraint to their prior values to maximize the likelihood~\cite{Collie}. We then use the profile likelihood ratio $L = -2\,\mathrm{ln}(P_{s+b}/P_b)$ as the test statistic, where $P_{s+b}$ is the Poisson likelihood to observe the data under the signal+background hypothesis and $P_b$ is the Poisson likelihood to observe the data under the background-only hypothesis. For the background-only hypothesis, we fit three components to the data. We constrain the $t\bar{t}$ production to its theoretical cross section, and the multijets background to the predicted number of events. For the other backgrounds, we constrain the relative fractions of the individual background sources to their predicted values and treat the overall normalization as a free parameter. For the signal+background fit we add $\sigma B$ for the resonance as a parameter to the fit. 

\begin{figure}[tb]
\epsfxsize=3.5in\epsfbox{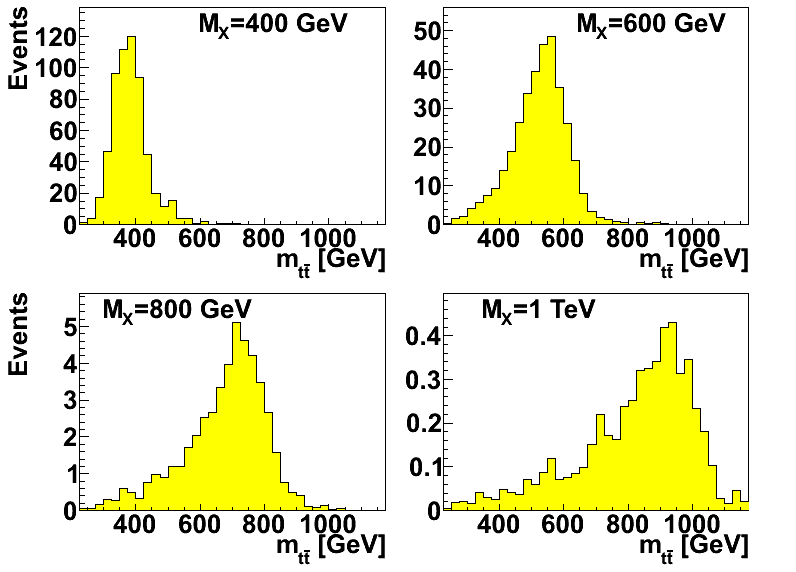}
\caption{Distributions of $m_{t\bar{t}}$ for resonances with mass $M_X$, normalized to the predicted $\sigma B$ from Table~\ref{tab:limits}.}
\label{fig:MCdist}
\end{figure}

We use the $CL_s$ method~\cite{CLs} to determine the limits on $\sigma B$. Using pseudoexperiments, we determine the probability to measure values of $L$ that are larger than the value observed in the data sample if there is a $t\bar{t}$ resonance signal, $CL_{s+b}$, and if there is no such signal, $CL_b$. The value of $\sigma B$ for which $1-CL_{s+b}/CL_b=0.95$ is the 95\% C.L. upper limit. We repeat this procedure at every resonance mass value. 

Table~\ref{tab:syst} summarizes the sources of systematic uncertainties in the normalizations of
the components of the model used in the limit calculation. No uncertainties are given for the physics 
backgrounds other than $t\bar t$ production because their normalizations are free parameters of the fit. When estimating the effect of uncertainties in the jet energy scale, the jet identification efficiency, and the jet energy resolution, we vary the shape of the $m_{t\bar{t}}$ distributions. We also assign an uncertainty to the shape of the $m_{t\bar{t}}$ distribution from $t\bar t$ production equal to the difference between the default simulation using {\sc alpgen} and {\sc pythia} and a simulation using {\sc mc@nlo} and {\sc herwig}. We also considered variations in the amount of initial and final state radiation but concluded that the resulting changes in the $m_{t\bar{t}}$ distribution were negligible relative to the uncertainties that we already consider. 

\begin{table}[tb]
\caption{\label{tab:syst} Summary of systematic uncertainties above 2\%. Some values vary with subsample. The numbers give the range of the uncertainties.}
\begin{ruledtabular}
\begin{tabular}{lccc}
Source                   & resonance & $t\bar{t}$ & multijets \\
\hline
$t\bar{t}$ cross section & ---          & 9\%        & ---\\
Multijets normalization  & ---          & ---        & (30--50)\% \\
Integrated luminosity    & 6.1\%        & 6.1\%      & ---\\
Monte Carlo model         & ---          & 4.3\%      & ---\\
Trigger efficiency       & $\leq$5\%    & $\leq$5\%  & ---\\
$b$-tagging efficiency   & (3--11)\%    & (3--5)\%   & ---\\
Lepton identification    & (3--4)\%     & (3--4)\%   & ---\\
Jet energy calibration   & (2--4)\%     & (2--5)\%   & ---\\
Jet energy resolution    & (3--5)\%     & (3--5)\%   & ---\\
Jet identification       & $\leq$7\%    & $\leq$10\% & ---\\
\end{tabular} 
\end{ruledtabular}
\end{table}

Figure~\ref{fig:data} shows the distribution of $m_{t\bar{t}}$ observed in data compared to expectations from standard model backgrounds and a 950~GeV $t\bar{t}$ resonance. We observe a small excess of events at high mass values. The excess is present in both the $e$+jets data and the $\mu$+jets data. We can fit the data best with an additional resonance signal with a mass of 950~GeV and $\sigma B=0.10\pm0.05$~pb. The value of $1-CL_b$ for the data gives the probability of getting a deviation of at least the observed size at this mass value from the standard model expectation in the absence of physics beyond the standard model. We find a $p$ value of 0.018, corresponding to 2.1 Gaussian-equivalent standard deviations. This significance value does not take into account that the excess may have occurred at any place in the mass spectrum.

\begin{figure}[tb]
\epsfxsize=3in\epsfbox{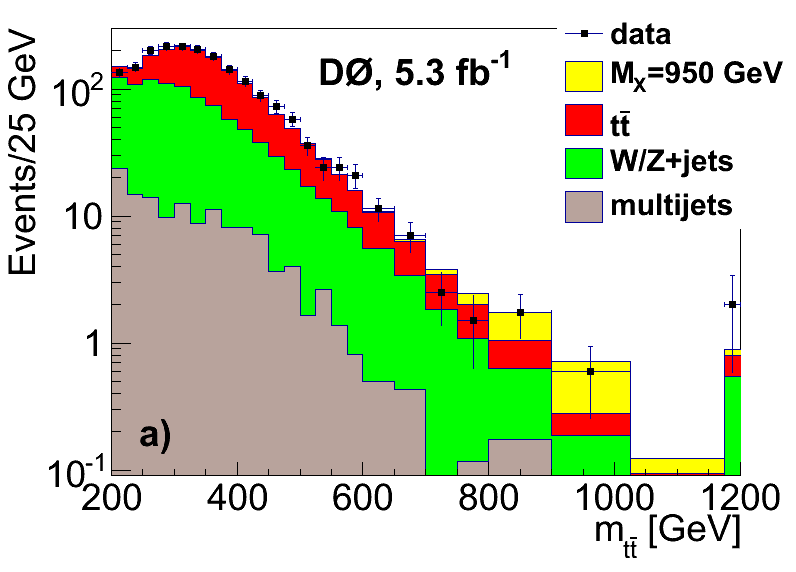}
\epsfxsize=3in\epsfbox{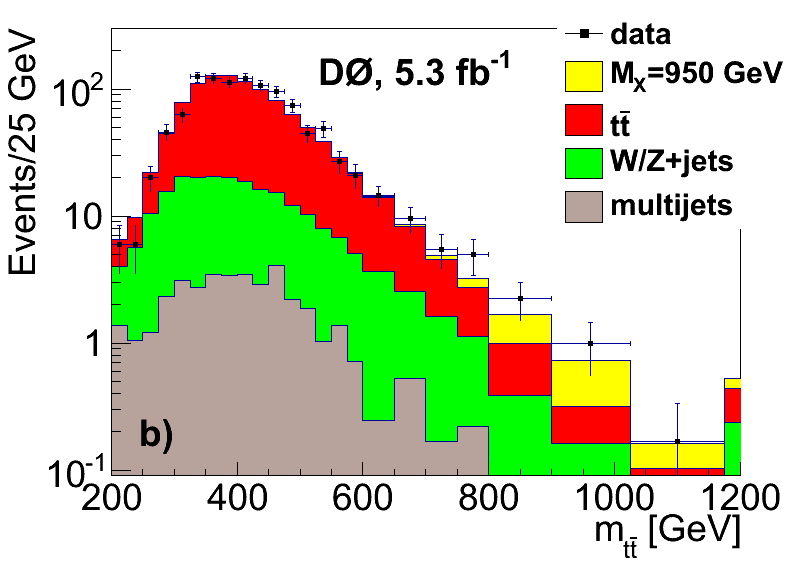}
\caption{Distribution of $m_{t\bar{t}}$ for events that pass the final event selection with (a) exactly 3 jets and (b) at least 4 jets, compared with expectations for standard model processes and a 950~GeV resonance signal with the best fitted $\sigma B=0.10$~pb. The highest bin in each histogram shows the  number of events with $m_{t\bar{t}}>1175$~GeV.
\label{fig:data}}
\end{figure}

Figure~\ref{fig:limits} and Table~\ref{tab:limits} display the resulting limits on $\sigma B$ compared to the limits expected in the absence of a narrow $t\bar{t}$ resonance and to the predicted NLO production cross section of a topcolor $Z'$ boson~\cite{Harris:1999ya} as a function of the resonance mass. The expected mass limit for production of such a $Z'$ boson that exclusively decays to $t\bar{t}$ is 920~GeV. The observed cross section limit excludes $Z'$ boson production at the 95\% C.L. for masses below 835~GeV. 

\begin{table}[tb]
\caption{\label{tab:limits} Observed and expected 95\% C.L. limits on $\sigma B$ compared with predictions for a topcolor $Z'$ computed assuming $m_t=172.5$~GeV and CTEQ6L1 parton distribution functions.}
\begin{ruledtabular}
\begin{tabular}{lccc}
$M_X$    & predicted $\sigma B$ & expected limit & observed limit \\
(GeV)    & (pb) & (pb) & (pb) \\
\hline
350   &     7.85~~      &     1.13 &    1.13 \\
400	&	12.79~~~ 	&	1.18	&	0.96	\\
450	&	8.59~~	&	0.92	&	1.16	\\
500	&	5.35~~	&	0.62	&	0.95	\\
550	&	3.32~~	&	0.42	&	0.39	\\
600	&	2.03~~	&	0.34	&	0.28	\\
650	&	1.24~~	&	0.26	&	0.19	\\
700	&	0.76~~	&	0.20	&	0.24	\\
750	&	0.46~~	&	0.16	&	0.16	\\
800	&	0.28~~	&	0.12	&	0.20	\\
850	&	0.17~~	&	0.10	&	0.19	\\
900	&	0.11~~	&	0.08	&	0.20	\\
950	&	0.059~	&	0.07	&	0.18	\\
1000	&	0.034~	&	0.07	&	0.16	\\
1050	&	0.020~	&	0.06	&	0.14	\\
1100	&	0.012~	&	0.06	&	0.14	\\
1150	&	0.0069	&	0.06	&	0.12	\\
1200	&	0.0041	&	0.07	&	0.13	\\
\end{tabular} 
\end{ruledtabular}
\end{table}

\begin{figure}[tb]
\epsfxsize=3in\epsfbox{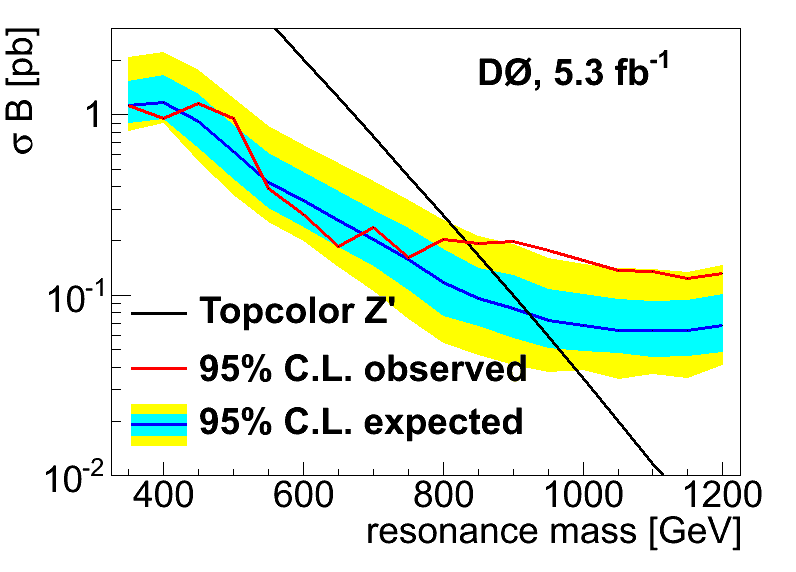}
\caption{Observed and expected upper limits on cross section times branching fraction $\sigma B$ for a narrow $t\bar{t}$ resonance as a function of the resonance mass. The shaded regions around the expected limit represent the $\pm$1 and $\pm$2 standard deviation bands. The solid line shows the predicted topcolor $Z'$ production cross section assuming $B(Z'\rightarrow t\bar{t})=100\%$. \label{fig:limits}}
\end{figure}

In our previous publication~\cite{Abazov:2008ny}, we computed a Bayesian limit on $\sigma B$ for a $t\bar{t}$ resonance. For reference, we have repeated the same calculation using a subset of the data sample reported in this paper, corresponding to $4.3\ifb$, and find  results consistent with the limits on $\sigma B$ given in Fig.~\ref{fig:limits}. We also find that our results are independent of the couplings of the $t\bar{t}$ resonance (pure vector, pure axial-vector, or standard-model $Z$-like) and are therefore valid for any narrow resonance decaying 100\% to a $t\bar t$ final state.

In conclusion, we searched for production of a narrow $t\bar{t}$ resonance in the lepton+jets channel. We do not observe a signal consistent with the production of such a resonance, although we observe a slight excess of events around 950~GeV. We set upper limits on the cross section times branching fraction for production of such a resonance for masses between 350 and 1200~GeV. We exclude at 95\% C.L. the production of a topcolor $Z'$ that decays exclusively to $t\bar{t}$ for mass values below 835~GeV.

\input{acknowledgement.tex}
\input{biblio}

\end{document}

%% file: author_list.tex
%
\affiliation{Universidad de Buenos Aires, Buenos Aires, Argentina}
\affiliation{LAFEX, Centro Brasileiro de Pesquisas F{\'\i}sicas, Rio de Janeiro, Brazil}
\affiliation{Universidade do Estado do Rio de Janeiro, Rio de Janeiro, Brazil}
\affiliation{Universidade Federal do ABC, Santo Andr\'e, Brazil}
\affiliation{Instituto de F\'{\i}sica Te\'orica, Universidade Estadual Paulista, S\~ao Paulo, Brazil}
\affiliation{University of Science and Technology of China, Hefei, People's Republic of China}
\affiliation{Universidad de los Andes, Bogot\'{a}, Colombia}
\affiliation{Charles University, Faculty of Mathematics and Physics, Center for Particle Physics, Prague, Czech Republic}
\affiliation{Czech Technical University in Prague, Prague, Czech Republic}
\affiliation{Center for Particle Physics, Institute of Physics, Academy of Sciences of the Czech Republic, Prague, Czech Republic}
\affiliation{Universidad San Francisco de Quito, Quito, Ecuador}
\affiliation{LPC, Universit\'e Blaise Pascal, CNRS/IN2P3, Clermont, France}
\affiliation{LPSC, Universit\'e Joseph Fourier Grenoble 1, CNRS/IN2P3, Institut National Polytechnique de Grenoble, Grenoble, France}
\affiliation{CPPM, Aix-Marseille Universit\'e, CNRS/IN2P3, Marseille, France}
\affiliation{LAL, Universit\'e Paris-Sud, CNRS/IN2P3, Orsay, France}
\affiliation{LPNHE, Universit\'es Paris VI and VII, CNRS/IN2P3, Paris, France}
\affiliation{CEA, Irfu, SPP, Saclay, France}
\affiliation{IPHC, Universit\'e de Strasbourg, CNRS/IN2P3, Strasbourg, France}
\affiliation{IPNL, Universit\'e Lyon 1, CNRS/IN2P3, Villeurbanne, France and Universit\'e de Lyon, Lyon, France}
\affiliation{III. Physikalisches Institut A, RWTH Aachen University, Aachen, Germany}
\affiliation{Physikalisches Institut, Universit{\"a}t Freiburg, Freiburg, Germany}
\affiliation{II. Physikalisches Institut, Georg-August-Universit{\"a}t G\"ottingen, G\"ottingen, Germany}
\affiliation{Institut f{\"u}r Physik, Universit{\"a}t Mainz, Mainz, Germany}
\affiliation{Ludwig-Maximilians-Universit{\"a}t M{\"u}nchen, M{\"u}nchen, Germany}
\affiliation{Fachbereich Physik, Bergische Universit{\"a}t Wuppertal, Wuppertal, Germany}
\affiliation{Panjab University, Chandigarh, India}
\affiliation{Delhi University, Delhi, India}
\affiliation{Tata Institute of Fundamental Research, Mumbai, India}
\affiliation{University College Dublin, Dublin, Ireland}
\affiliation{Korea Detector Laboratory, Korea University, Seoul, Korea}
\affiliation{CINVESTAV, Mexico City, Mexico}
\affiliation{Nikhef, Science Park, Amsterdam, the Netherlands}
\affiliation{Radboud University Nijmegen, Nijmegen, the Netherlands and Nikhef, Science Park, Amsterdam, the Netherlands}
\affiliation{Joint Institute for Nuclear Research, Dubna, Russia}
\affiliation{Institute for Theoretical and Experimental Physics, Moscow, Russia}
\affiliation{Moscow State University, Moscow, Russia}
\affiliation{Institute for High Energy Physics, Protvino, Russia}
\affiliation{Petersburg Nuclear Physics Institute, St. Petersburg, Russia}
\affiliation{Instituci\'{o} Catalana de Recerca i Estudis Avan\c{c}ats (ICREA) and Institut de F\'{i}sica d'Altes Energies (IFAE), Barcelona, Spain}
\affiliation{Stockholm University, Stockholm and Uppsala University, Uppsala, Sweden}
\affiliation{Lancaster University, Lancaster LA1 4YB, United Kingdom}
\affiliation{Imperial College London, London SW7 2AZ, United Kingdom}
\affiliation{The University of Manchester, Manchester M13 9PL, United Kingdom}
\affiliation{University of Arizona, Tucson, Arizona 85721, USA}
\affiliation{University of California Riverside, Riverside, California 92521, USA}
\affiliation{Florida State University, Tallahassee, Florida 32306, USA}
\affiliation{Fermi National Accelerator Laboratory, Batavia, Illinois 60510, USA}
\affiliation{University of Illinois at Chicago, Chicago, Illinois 60607, USA}
\affiliation{Northern Illinois University, DeKalb, Illinois 60115, USA}
\affiliation{Northwestern University, Evanston, Illinois 60208, USA}
\affiliation{Indiana University, Bloomington, Indiana 47405, USA}
\affiliation{Purdue University Calumet, Hammond, Indiana 46323, USA}
\affiliation{University of Notre Dame, Notre Dame, Indiana 46556, USA}
\affiliation{Iowa State University, Ames, Iowa 50011, USA}
\affiliation{University of Kansas, Lawrence, Kansas 66045, USA}
\affiliation{Kansas State University, Manhattan, Kansas 66506, USA}
\affiliation{Louisiana Tech University, Ruston, Louisiana 71272, USA}
\affiliation{Boston University, Boston, Massachusetts 02215, USA}
\affiliation{Northeastern University, Boston, Massachusetts 02115, USA}
\affiliation{University of Michigan, Ann Arbor, Michigan 48109, USA}
\affiliation{Michigan State University, East Lansing, Michigan 48824, USA}
\affiliation{University of Mississippi, University, Mississippi 38677, USA}
\affiliation{University of Nebraska, Lincoln, Nebraska 68588, USA}
\affiliation{Rutgers University, Piscataway, New Jersey 08855, USA}
\affiliation{Princeton University, Princeton, New Jersey 08544, USA}
\affiliation{State University of New York, Buffalo, New York 14260, USA}
\affiliation{Columbia University, New York, New York 10027, USA}
\affiliation{University of Rochester, Rochester, New York 14627, USA}
\affiliation{State University of New York, Stony Brook, New York 11794, USA}
\affiliation{Brookhaven National Laboratory, Upton, New York 11973, USA}
\affiliation{Langston University, Langston, Oklahoma 73050, USA}
\affiliation{University of Oklahoma, Norman, Oklahoma 73019, USA}
\affiliation{Oklahoma State University, Stillwater, Oklahoma 74078, USA}
\affiliation{Brown University, Providence, Rhode Island 02912, USA}
\affiliation{University of Texas, Arlington, Texas 76019, USA}
\affiliation{Southern Methodist University, Dallas, Texas 75275, USA}
\affiliation{Rice University, Houston, Texas 77005, USA}
\affiliation{University of Virginia, Charlottesville, Virginia 22901, USA}
\affiliation{University of Washington, Seattle, Washington 98195, USA}
\author{V.M.~Abazov} \affiliation{Joint Institute for Nuclear Research, Dubna, Russia}
\author{B.~Abbott} \affiliation{University of Oklahoma, Norman, Oklahoma 73019, USA}
\author{B.S.~Acharya} \affiliation{Tata Institute of Fundamental Research, Mumbai, India}
\author{M.~Adams} \affiliation{University of Illinois at Chicago, Chicago, Illinois 60607, USA}
\author{T.~Adams} \affiliation{Florida State University, Tallahassee, Florida 32306, USA}
\author{G.D.~Alexeev} \affiliation{Joint Institute for Nuclear Research, Dubna, Russia}
\author{G.~Alkhazov} \affiliation{Petersburg Nuclear Physics Institute, St. Petersburg, Russia}
\author{A.~Alton$^{a}$} \affiliation{University of Michigan, Ann Arbor, Michigan 48109, USA}
\author{G.~Alverson} \affiliation{Northeastern University, Boston, Massachusetts 02115, USA}
\author{G.A.~Alves} \affiliation{LAFEX, Centro Brasileiro de Pesquisas F{\'\i}sicas, Rio de Janeiro, Brazil}
% ...
%
%
\collaboration{The D0 Collaboration\footnote{with visitors from
$^{a}$Augustana College, Sioux Falls, SD, USA,
$^{b}$The University of Liverpool, Liverpool, UK,
$^{c}$UPIITA-IPN, Mexico City, Mexico,
$^{d}$SLAC, Menlo Park, CA, USA,
$^{e}$University College London, London, UK,
$^{f}$Centro de Investigacion en Computacion - IPN, Mexico City, Mexico,
$^{g}$ECFM, Universidad Autonoma de Sinaloa, Culiac\'an, Mexico,
and 
$^{h}$Universit{\"a}t Bern, Bern, Switzerland.
$^{\ddag}$Deceased.
}} \noaffiliation
\vskip 0.25cm

%% file: acknowledgement.tex
%
We thank the staffs at Fermilab and collaborating institutions,
and acknowledge support from the
DOE and NSF (USA);
CEA and CNRS/IN2P3 (France);
FASI, Rosatom and RFBR (Russia);
CNPq, FAPERJ, FAPESP and FUNDUNESP (Brazil);
DAE and DST (India);
Colciencias (Colombia);
CONACyT (Mexico);
KRF and KOSEF (Korea);
CONICET and UBACyT (Argentina);
FOM (The Netherlands);
STFC and the Royal Society (United Kingdom);
MSMT and GACR (Czech Republic);
CRC Program and NSERC (Canada);
BMBF and DFG (Germany);
SFI (Ireland);
The Swedish Research Council (Sweden);
and
CAS and CNSF (China).

%% file: biblio.tex
\providecommand{\href}[2]{#2}
\begingroup\raggedright\endgroup